\documentstyle{elsart}
\begin{document}
\begin{frontmatter}
\title{Cuprate interband model and doping dependence of the
coherence length}
\author[physics,kolja]{N.Kristoffel\thanksref{mail},}
\author[kolja]{T.\"Ord,}
\author[physics]{P.Rubin}
\address[physics]{Institute of Physics, University of Tartu,
Riia 142, 51014 Tartu, Estonia}
\address[kolja]{Institute of Theoretical Physics, University of Tartu, T\"ahe 4,
51010 Tartu, Estonia}
\thanks[mail]{Corresponding author: Tel: +372 742 8164; fax: +372 738 3033;
E-mail address: kolja@fi.tartu.ee}

\begin{abstract}
The free energy expansion of a two-band pair-transfer
superconductor is developed. A critical in temperature and a
noncritical coherence length appear. The effective in-plane
$\xi_{ab}$, $H_{c2}\sp c\sim \xi_{ab}\sp{-2}$ and the
thermodynamic critical magnetic field ($H_{c0}$) on the whole hole
doping ($p$) scale are calculated for a "typical'' cuprate. A
doping-prepared bare spectrum with normal state gaps quenched by
doping has been used. The coherence length $\xi_{ab}$ falls with
$p$ first rapidly. A moderate enhancement starts when the "cold''
defect band overlap with the valence band is reached. At
overdoping $\xi_{ab}$ rises markedly. The $H_{c2}\sp c$ curve
shows a maximum before $T_{cmax}$. The bell-like $H_{c0}$-curve
follows closely $T_c$, superconducting gaps, and superfluid
density. The theoretical $\xi_{ab}(p)$ nonmonotonic curve on the
whole doping scale agrees with a recent experimental finding. \\

\end{abstract}
\end{frontmatter}

\section{Introduction}
Two-band superconductivity with the interband pairing channel
[1,2] is known for a long time. It is seemingly the most effective
mechanism which is able to serve high transition temperatures
($T_c$) in a simple way. The interest for this mechanism and its
applications has recently essentially grown in connection with the
now recognized two-gap superconductivity, e.g. [3-5], of magnesium
diboride. Various two-band approaches with diverse electron
spectra have been applied also for cuprates (e.g. introducting
reviews [6,7]). Wide experimental data on cuprate energetic
characteristics ($T_c$, superconducting-, normal state- and
pseudogaps) and their interrelations have been ordered quite
recently. Data characterizing the coherence properties are far
from being complete. Cuprate superconductivity mechanism itself
remains debatable.

In papers [8-11] a simple, partly postulative model which uses
only very general knowledge on cuprates has been proposed. The
electron spectrum of a doped cuprate incorporates in this model
the valence band and hole-created defect states ("hot'' and
"cold'' subbands [11]) with bare normal state gaps between them.
These gaps become quenched on progressive doping. Overlap dynamics
of bands appears as a novel source of critical doping
concentrations. The leading pairing interaction is supposed to be
the pair-transfer between the itinerant and defect subsystem
states. The nature of the minimal quasiparticle excitation energy
changes on doping. This model is able to describe the observed
behaviour of cuprate energetic characteristics on the doping
scale. The pseudogap(s) are of "extrinsic'' occurrence and
transform as precursors on the doping scale (not the energetic
scale) to superconducting gaps. Pseudogaps survive as normal state
gaps. These latter do not manifest in the superconducting density
[12] because of interband nature of the pairing. Superfluid
density follows the trends shown by $T_c$ and superconducting gaps
with expressed maximum near the optimal doping. The strength of
the pairing and the phase coherence develop and vanish
simultaneously as recent experimental findings [13-18] stress.

Experimental data on cuprate characteristics reflecting the
coherence properties (coherence lengths and critical magnetic
fields) are given usually for dispersive dopings and temperatures
(e.g. [6,19]). The behaviour for extended dopings have been
obtained in few cases, and a common trend has not been formulated.
For YBa$_2$Cu$_3$O$_y$ as a general trend the decrease of the
in-plane coherence length $\xi_{ab}(0)$ with increased doping has
been found until the $T_c$ maximum is reached [20]. It manifests
also in $H_{c2}\sp c$ enhancement with $y$. For
La$_{2-x}$Sr$_x$CuO$_4$ the c-axis coherence length amplitude
diminishes slowly with $x$ and falls then to a remarkably smaller
nearly constant value on overdoping [21]. On the contrary, for the
Bi-based systems the opposite behaviour has been found in [22,23].
The pair coherence length rises here with doping in a wide region.
The corresponding $H_{c2}$ falls off. The authors of [22,23]
stress the pseudogap to represent the amplitude of the pairing
strength. This is not the case in our model [8-12]. The
antiferromagnetic coherence length representing the average
distance between the doped holes falls off with enhanced doping as
$\sim p\sp{-1/2}$ [24].

However, in the recent investigation [25] based on the vortex core
size measurements, seemingly for the first time, the doping
dependence of the cuprate coherence length on the whole doping
scale has been obtained from the low temperature data.

In the present communication we apply the model of [11,12] to
investigate the in-plane coherence length amplitude at zero
temperature $\xi_{ab}(0)= \xi_0$ and the corresponding $H_{c2}\sp
c$, and also the thermodynamic critical field $H_c(0)$, on the
full doping scale for a "typical'' cuprate. Qualitative agreement
with the valley-profile type curve recently found in [25] is
obtained.

We start with the derivation of the long-wavelength expansion of
the free energy for a two-band superconductor with leading
pair-transfer interaction.

\section{The free energy}
We describe the superconducting system under consideration by the
Hamilton operators separation as
\begin{equation}
    H=H_1+H_2, \; \; \; H_1=H_0+H_{\lambda}, \; \; \;
    H_2=-H_{\lambda}+H_i
\end{equation}
with
\begin{equation}
H_0=\sum_{\sigma \vec{k}s}\tilde{\epsilon}_{\sigma}(\vec{k}) a\sp
+_{\sigma \vec{k}s}a_{\sigma \vec{k}s}\; ,
\end{equation}
\begin{equation}
H_i =2W \sum_{\sigma ,\sigma
'}'\sum_{\vec{k},\vec{k}'}\sum_{\vec{q}}a\sp
+_{\sigma\vec{k}\uparrow} a\sp +_{\sigma
-\vec{k}+\vec{q}\downarrow}a_{\sigma '
-\vec{k}'+\vec{q}'\downarrow} a_{\sigma ' \vec{k}'\uparrow}\; ,
\end{equation}

\begin{equation}
H_{\lambda} =2W \sum_{\sigma \vec{k}\vec{q}} (\lambda_{\sigma
\vec{q}}a\sp +_{\sigma\vec{k}\uparrow} a\sp +_{\sigma
-\vec{k}\downarrow} + \rm{h.c.}) \; .
\end{equation}
Here $\tilde{\epsilon}_{\sigma}=\epsilon_{\sigma}-\mu$, $\mu$ is
the chemical potential, $\sigma =1, 2$ is the electron subsystem
index, $s$ is the spin index, $W$ is the interband pair-transfer
interaction constant, $\vec{q}$ stands for the total momentum of
an electron pair. Correspondingly the superconducting order
parameters $\overline{\lambda}_{\sigma\vec{q}}$ can express
spatially inhomogeneous fluctuations.

The mean-field free energy of $H$ reads
\begin{equation}
F=-k_BT +\ln Z(H_1)+\langle H_2\rangle_{H_1}\; ,
\end{equation}
$$
Z(H_1)=Sp \exp \left(-\frac{H_1}{k_BT}\right)\; ,
$$
and  $\langle ...\rangle_{H_1}$ means the average in respect of
$H_1$. Near the phase transition temperature (5) will be expanded
as
\begin{equation}
F=F_0+F_2+F_4\; ,
\end{equation}
where $F_0=-k_BT \ln Z(H_0)$, $F_2\sim \lambda\sp 2$, $F_4\sim
\lambda\sp 4$. For the long-wave fluctuations of $\lambda$, $F_4$
can be taken as in the case of $\vec{q}=0$ [26]. Application of
the unitary transformation $U=\exp (iS)$ with
\begin{equation}
S=2Wi\sum_{\sigma\vec{k}\vec{q}} \left[\frac{a\sp
+_{\sigma\vec{k}\uparrow} a\sp +_{\sigma -\vec{k}\downarrow}}
{\tilde{\epsilon}_{\sigma}(\vec{k})+\tilde{\epsilon}_{\sigma}(\vec{k}-\vec{q})}
\lambda_{\sigma\vec{q}}+\rm{h.c.}\right]
\end{equation}
enables to calculate the terms in (5) of the necessary order. One
ends with the expression
\begin{eqnarray}
F_2=|W|\sum_{\vec{q}}\left[\sum_{\sigma}\eta_{\sigma\vec{q}}(T)
|\lambda_{\sigma\vec{q}}|\sp 2 \right. \nonumber \\
\left. + \frac{w}{2}\eta_{1\vec{q}}(T)\eta_{2\vec{q}}(T)
\left(\lambda_{1\vec{q}}\lambda_{2\vec{q}}\sp *+
\lambda_{1\vec{q}}\sp *\lambda_{2\vec{q}}\right)\right]\; ,
\end{eqnarray}
 where $w=sgn(W)$ and
\begin{eqnarray}
\eta_{\sigma\vec{q}}(T)=2|W|\sum_{\vec{k}}\left[
\tilde{\epsilon}_{\sigma}(\vec{k})+
\tilde{\epsilon}_{\sigma}(\vec{k}-\vec{q})\right]\sp{-1} \nonumber
\\ \times
\left[ {\rm th} \frac{\tilde{\epsilon}_{\sigma}(\vec{k})}{2k_BT}
+{\rm th}
\frac{\tilde{\epsilon}_{\sigma}(\vec{k}-\vec{q})}{2k_BT}\right]\;
.
\end{eqnarray}

\section{Expansions and equilibrium equations}
The quantity (9) will be expressed near $T=T_c$ ($\tau
=(T-T_c)T_c\sp{-1}$) and $\vec{q}=0$ as
\begin{equation}
\eta_{\sigma\vec{q}}(T)=\eta_{\sigma
0}(\theta_c)-\alpha_{\sigma}\tau- \sum_{i,j}B_{\sigma ij}q_iq_j\;
,
\end{equation}
where ($\theta_c=k_BT_c$)
\begin{equation}
\alpha_{\sigma}=\frac{|W|}{\theta_c}\sum_{\vec{k}}ch\sp{-2}
\frac{\tilde{\epsilon}_{\sigma}(\vec{k})}{2\theta_c}
\end{equation}
and $B_{\sigma}$ is given by a complicated formula for which an
approximate expression will be given later. The linear terms in
(10) are absent in connection with the approximation
\begin{equation}
\epsilon_{\sigma}(\vec{k})=E_{0\sigma}\pm \sum_i\frac{\hbar\sp
2k_i\sp 2} {2m_{\sigma i}}
\end{equation}
for the "band'' energies with $m_{\sigma i}>0$.

The superconducting phase transition temperature $T_c$ is
determined by the equation [7]
\begin{equation}
\eta_1\eta_2=4
\end{equation}
with $\eta_{0\sigma}(\theta_c)=\eta_{\sigma}$.

The integration in the momentum space is supposed to be performed
from $\Gamma_{0\sigma}$ to $\Gamma_{c\sigma}$ for an electron band
and from $\Gamma_{c\sigma}$ to $\Gamma_{0\sigma}$ for a hole band.
For constant densities of states ($\rho_{\alpha}$) having in mind
CuO$_2$ planes and supposing that $|\Gamma -\mu|>\theta_c$, one
obtains in the case where $\mu$ lies inside of the integration
limits
\begin{equation}
\eta_{\sigma}=2|W|\rho_{\sigma}\ln |\Gamma_{0\sigma}-\mu
||\Gamma_{c\sigma}-\mu |
\left(\frac{2\gamma}{\pi\theta_c}\right)\sp 2 \; ,
\end{equation}
\begin{equation}
\alpha_{\sigma}=4|W|\rho_{\sigma}\; ,
\end{equation}
\begin{equation}
\sum_{i,j}B_{\sigma ij}q_iq_j=\beta_{\sigma}\sum_i \frac{\hbar\sp
2q_i\sp 2}{4m_{\sigma i}}\; ,
\end{equation}
\begin{equation}
\beta_{\sigma}=\frac{7 \zeta (3)|W|\rho_{\sigma}|\mu
-E_{0\sigma}|} {(\pi\theta_c)\sp 2}
\end{equation}
with $\gamma\approx 1.78$ and $\zeta (3)\approx 1.2$. Note that
$|\mu -E_{0\sigma}|$ determines the Fermi energy in the
corresponding band. In the case if $\mu$ lies out of the
integration limits
\begin{equation}
\eta_{\sigma}=2|W|\rho_{\sigma}\ln
\left|\frac{\Gamma_{c\sigma}-\mu} {\Gamma_{0\sigma}-\mu}\right|\;
, \; \; \; \; \alpha_{\sigma}=\beta_{\alpha}=0\; .
\end{equation}
Analogous formula can be found also for $\mu$ coincidence with one
of the integration limits.

After the Fourier transformation to the coordinate space the
approximate expression for the second order contribution to the
free energy (8) reads
$$
F_2=\frac{|W|}{V}\int d\vec{r}\left\{
\sum_{\sigma}\Bigg[(\eta_{\sigma}-
\alpha_{\sigma}\tau)|\lambda_{\sigma}(\vec{r})|\sp 2 \right.
$$
$$
\left. -\beta_{\sigma} \sum_i\frac{\hbar\sp 2}{4m_{\sigma
i}}\left|\left(\nabla_i-\frac{2ie}{\hbar c}
A_i\right)\lambda_{\sigma}(\vec{r})\right|\sp 2 \right]
$$
$$
+\frac{w}{2}\bigg[ [4-(\eta_1\alpha_2+\eta_2\alpha_1)\tau ]
(\lambda_1(\vec{r})\lambda_2\sp *(\vec{r})+\lambda_1\sp *(\vec{r})
\lambda_2(\vec{r}))
$$
$$
-\sum_i\left(\eta_1\beta_2\frac{\hbar}{4m_{2i}}+
\eta_2\beta_1\frac{\hbar}{4m_{1i}}\right)  $$
$$
\times \left[(\nabla_i-\frac{2ie}{\hbar c}A_i)\lambda_1(\vec{r})
\left((\nabla_i-\frac{2ie}{\hbar
c}A_i)\lambda_2(\vec{r})\right)\sp *\right.
$$
$$
 \left.\left.+\left((\nabla_i-\frac{2ie}{\hbar
c}A_i)\lambda_1(\vec{r})\right)\sp * (\nabla_i-\frac{2ie}{\hbar
c}A_i)\lambda_2(\vec{r})\right] \right]
$$
\begin{equation} \left.
+\frac{(rot \vec{A})\sp
2}{8\pi}\right\}\;
\end{equation} ,
 where the vector potential
$\vec{A}$ takes care for the gauge invariance. This $F_2$
expression contains squared gradient terms, however, with opposite
signs in comparison with the one band case. Products of gradient
terms of different bands appear in (19).

For the fourth order term analogously to [27]
$$
F_4=-\frac{|W|}{V}\int d\vec{r}\left\{
\frac{3}{2}\sum_{\sigma}\nu_{\sigma}
|\lambda_{\sigma}(\vec{r})|\sp 4+\right.
$$
$$
%\begin{equation}
+\frac{w}{2}[\nu_1\eta_2|\lambda_1(\vec{r})|\sp 2+
\nu_2\eta_1|\lambda_2(\vec{r})|\sp 2]
$$
\begin{equation}
\left. [\lambda_1(\vec{r}) \lambda_2(\vec{r})\sp
*+\lambda_1(\vec{r})\sp *\lambda_2(\vec{r})]\right\} d\vec{r}\; ,
\end{equation}
with
\begin{equation}
\nu_{\sigma}=\frac{14\zeta (3)|W|\sp
3\rho_{\sigma}}{(\pi\theta_c)\sp 2}\; ,
\end{equation}
and $\nu_{\sigma}=0$ for $\mu$ out of the band.

\section{Fluctuative ordering}
The expression (19) for $F_2$ can be diagonalized by an orthogonal
transformation
\begin{eqnarray}
\lambda_{1\vec{q}} & = & \lambda_{r\vec{q}}\cos \varphi_{\vec{q}}+
\lambda_{s\vec{q}}\sin \varphi_{\vec{q}}\nonumber \\
\lambda_{2\vec{q}} & = & -\lambda_{r\vec{q}}\sin
\varphi_{\vec{q}}+ \lambda_{s\vec{q}}\cos \varphi_{\vec{q}}\; ,
\end{eqnarray}
\begin{eqnarray}
\tan \varphi_{\vec{q}} =  \frac{w\eta_2}{2} \bigg[
\frac{(\eta_1\sp 2\alpha_2-\eta_2\sp 2\alpha_1)\tau
 +  \sum_i \bigg(\frac{\eta_1\sp 2\beta_2}{m_{2i}}-\frac{\eta_2\sp
2\beta_1}{m_{1i}} \bigg) \frac{\hbar\sp 2q_i\sp 2}{4}
}{4(\eta_1+\eta_2)} -1 \bigg]
\end{eqnarray}
to the new variables
\begin{equation}
\psi_{s,r}(\vec{r})=\sqrt{\frac{|W|(\beta_1+\beta_2)}{2V}}\sum_{\vec{q}}
\lambda_{s,r}(\vec{q})e\sp{i\vec{q}\vec{r}}\; .
\end{equation}
Then
$$
F_2=\int d\vec{r} \bigg[ a_s\tau |\psi_s(\vec{r})|\sp 2
+(a_o-a_r\tau )|\psi_r(\vec{r})|\sp 2
$$
$$
+\sum_i \frac{\hbar\sp
2}{4M_{si}}\left|\left(\nabla_i-\frac{2ie}{\hbar c}A_i\right)
\psi_s(\vec{r})\right|\sp 2
$$
\begin{equation}  -\sum_i \frac{\hbar\sp
2}{4M_{ri}}\left|\left(\nabla_i-\frac{2ie}{\hbar c}A_i\right)
\psi_r(\vec{r})\right|\sp 2 + \frac{(rot\vec{A})\sp 2}{8\pi}
\bigg] \; .
\end{equation}
Here
\begin{equation}
a_0=\frac{(\eta_1+\eta_2)\sp 2}{u}\;, \; \; \;
u=\frac{1}{2}(\eta_1+\eta_2)(\beta_1+\beta_2)\; ,
\end{equation}
\begin{equation}
a_s=\frac{\eta_1\alpha_2+\eta_2\alpha_1}{u}\; , \; \; \;
a_r=\frac{(2\eta_1+\eta_2)\alpha_2+(2\eta_2+\eta_1)\alpha_1}{u}\;
,
\end{equation}
$$
M_{si}\sp{-1}=\frac{\eta_1\beta_2m_{2i}\sp{-1}+\eta_2\beta_1m_{1i}\sp{-1}}
{u}\; , \; \;
$$
\begin{equation}
M_{ri}\sp{-1}=\frac{(2\eta_1+\eta_2)\beta_2m_{2i}\sp{-1}+
(2\eta_2+\eta_1)\beta_1m_{1i}\sp{-1}}{u}\; .
\end{equation}

The indices $s$ and $r$ correspond here to "soft'' and "rigid''.
The coefficient $a_s\tau$ changes its sign when temperature passes
$T_c$. On the contrary, the coefficient before $|\psi_r|\sp 2$
remains positive. The "soft'' variable $\psi_s$ plays the role of
a driver for the phase transition. Indeed, proceeding from the
system for the order parameters minimizing the free energy
\begin{eqnarray}
2w\overline{\lambda}_1+[\eta_2-\sum_i\frac{\beta_2}{4m_{2i}}
(-i\hbar\nabla_i-\frac{2e}{c}A_i)\sp 2 \nonumber \\
-\alpha_2\tau -
\nu_2|\overline{\lambda}_2|\sp 2]\overline{\lambda}_2 & = & 0 \nonumber \\
2w\overline{\lambda}_2+[\eta_1-\sum_i\frac{\beta_1}{4m_{1i}}
(-i\hbar\nabla_i-\frac{2e}{c}A_i)\sp 2- \nonumber \\
 \alpha_1\tau
- \nu_1|\overline{\lambda}_1|\sp 2]\overline{\lambda}_1 & = & 0\;
,
\end{eqnarray}
for the homogeneous case ($\vec{A}=0$ and $T<T_c$), one finds
\begin{equation}
|\overline{\lambda}_{1,2}|\sp 2=-\eta_{2,1}\Xi\tau \; ,
\end{equation}
\begin{equation}
\Xi_2 =\frac{\eta_1\alpha_2+\eta_2\alpha_1}{\eta_1\sp 2\nu_2+
\eta_2\sp 2\nu_1}\; .
\end{equation}

The superconducting gaps $\Delta_{1,2}=2W\overline{\lambda}_{1,2}$
vanish simultaneously at $T_c$. According to (24)
\begin{equation}
|\overline{\psi}_s|\sp 2=-\frac{|W|}{V}u\Xi\tau \; ,
\end{equation}
$$
|\overline{\psi}_r|\sp 2=0 \; ,
$$
and for $T>T_c$, $\overline{\psi}_s=\overline{\psi}_r=0$. The
superconducting gaps are determined only by the "soft" order
parameter.

Minimizing (25) for $\vec{A}=0$ in the normal phase shows that
fluctuations of $\psi_s$ and $\psi_r$ satisfy the equations
\begin{equation}
\sum_i\xi_{s,r,i}\sp 2\nabla_i\sp 2\psi_{s,r}=\psi_{s,r}\; ,
\end{equation}
where the corresponding coherence lengths are given by
\begin{equation}
\xi_{si}\sp 2=\frac{\hbar\sp 2}{4M_{si}a_s\tau}\; ,
\end{equation}
\begin{equation}
\xi_{ri}\sp 2=-\frac{\hbar\sp 2}{4M_{ri}(a_0-a_r\tau)}\; ,
\end{equation}

As the result, the fluctuations of band order parameters are
governed by two special coherence scales. The $\xi_s$ acts as an
Ornstein-Zernicke type critical coherence length known for
one-band systems and diverging at $T\rightarrow T_c$. The other,
$\xi_r$, behaves noncritically (is rigid) and as an imaginary
quantity characterizes a periodic spatial coherence wave. In the
limit $\tau \rightarrow 0$, $\xi_s\rightarrow \infty$, however
$\xi_r$ remains finite. Analogously a two-band superconductor with
interband pairing possesses a critical and a noncritical
relaxation channel [27]. The relaxation times connected with these
channels show correspondingly a critical and noncritical
dependence on temperature.

The discussion of the nature of the spatial periodic fluctuating
structure characterized by the rigid coherence length $\xi_r$ with
weak dependence on temperature remains out of the scope of the
present paper. However, it will be tempting to attribute $\xi_r$
to the recently observed pair density wave in the Bi-compound
[28]. Then seemingly $\xi_r$ characterizes the average distance
between the pairs (without phase coherence in the normal state).
As a function of doping $\xi_r$ falls off with leaving the
underdoped region and remains nearly constant in the basic region
of actual $T_c$-s. Then the pair distribution density becomes more
homogeneous, cf. [29]. Note that a spatial modulation of gap sizes
has also been observed [30].

\section{Cuprate superconductivity coherence and critical magnetic fields}
In general the cuprate superconductivity two-component scenario
supposed in [8-11] supports on the separation of the strongly
correlated doped CuO$_2$ planes in nanoscale structural
components. A new supercondoctivity playground is built up with
radical reorganization of the electron spectrum until the creation
of states corresponding to the doped-hole bearing subsystem. This
spectrum is doping-variable in band structure and densities. A new
pairing channel with interband pair transfer opens between the
(mainly) itinerant and defect components.

In what follows the cuprate critical coherence length $\xi_s
=\xi_0 \tau\sp{-1/2}$ will be calculated using the model of [11].
For the valence band $\Gamma_{c\gamma}=-D$, $\Gamma_{0\gamma}=0$
and for the defect system subbands $\Gamma_{c\alpha}=d_1-\alpha
c$, $\Gamma_{0\alpha}=d_1$ and $\Gamma_{c\beta}=d_2-\beta c$,
$\Gamma_{0\beta}=d_2$ ($d_1>d_2$, $\alpha >\beta$). The measure of
the doped hole concentration $c$ is scaled to the transition
temperature maximum at $p =0.16$ by $p =0.28c$. The bottoms of the
defect subbands (attributed to ($\frac{\pi}{2},\frac{\pi}{2}$) and
($\pi,0$) - type regions of the momentum space) evolve down in
energy with doping. The overlap with the itinerant band is reached
correspondingly at $c_{\beta} = d_2/\beta$ and $c_{\alpha} =
d_1/\alpha$. There are the following different arrangements of the
bands and $\mu$. At $c<c_{\beta}$ $\mu_1=d_2-\beta c$ remains
connected with the "cold'' $\beta$-band. In this very underdoped
region the charge carriers concentrate first. For $c>c_{\beta}$
$\mu_2=(d_2-\beta c)(1+2(1-c)\beta D\sp{-1})\sp{-1}$ is shifted
into the valence band. The overlap of the narrow $\beta$ band with
the wide $\gamma$ band leads to the formation of two Fermi surface
sheets with a tendency to appearance of a "flat band'' component
with lowering $\mu$. For the effective doping near $c_0$, defined
by $d_1-\alpha c_0=\mu_2$, the role of the "hot'' region is
enhanced. The $\mu_3=[\alpha d_2+\beta d_1- 2\alpha\beta c][\alpha
+\beta +(1-c)2\alpha\beta D\sp{-1}]\sp{-1}$ intersects all three
overlapping bands and $T_c$ becomes maximized. For the extended
overdoping $c>c_1$, $\mu_3=d_2-\beta c_1$, the chemical potential
falls out of the defect $\beta$-band. In the doping process the
mixing of the band components stimulates the Fermi-liquid
behaviour of the carriers. The decrease of $T_c$ at overdoping is
connected with the deterioration of the interband pairing
conditions. This leads to the drop of superfluid density on the
background of extending hole concentrations.

One finds in this model ($\xi_s^2=\xi_0^2 \tau^{-1}$)
\begin{equation}
\xi_0\sp 2=\frac{\hbar^2}{4}\frac{(\eta_{\alpha}+\eta_{\beta})
\beta_{\gamma}m_{\gamma}\sp{-1}+\eta_{\gamma}(\beta_{\alpha}m_{\alpha}\sp{-1}+
\beta_{\beta}m_{\beta}\sp{-1})}{(\eta_{\alpha}+\eta_{\beta})\alpha_{\gamma}+
\eta_{\gamma}(\alpha_{\alpha}+\alpha_{\beta})}\; ,
\end{equation}
which must be interpreted as the in-plane coherence length
($\xi_{ab}$).

The effective masses of the bands are connected with the constant
densities of states $m_i\sp{-1}=\frac{V}{2\pi\hbar\sp 2\rho_i}$,
$\rho_{\alpha}=1/2\alpha$, $\rho_{\beta}=1/2\beta$,
$\rho_{\gamma}=(1-c)/D$; $V$ is the plaquette area ($a\sp 2$) in
the CuO$_2$ plane.

The calculated $\xi_0$ over the whole hole doping region of a
"typical'' cuprate together with the $T_c$-curve is given in
Fig.1. The same plausible parameter set as  in [11] has been taken
for the illustration. The order of $\xi_0$ of some tens of \AA$\;$
in the actual region agrees with  the values given in the
literature for cuprates [6]. This points to some self-consistency
of the theoretical scheme concerning the earlier results for
energetic characteristics [8-12].
%\begin{figure}
%\resizebox{0.5\columnwidth}{!}

%\vspace{5cm}       % Give the correct figure height in cm
%\caption{Cuprate transition temperature (dashed line) and the
%ab-plane coherence length vs hole doping.}
%\end{figure}

The general trend of $\xi_0(p)$ found starts with the steep
decrease until the critical point $c_{\beta}$ is reached. Then a
moderate enhancement until the second critical point $c_0$ with
further quick increase follows. The behaviour of $\xi_0$ as
projected  on the $T_c$ bell-like curve wings seems to be a
natural result. At extreme dopings $\xi_0$ diverges.

The theoretical valley-profile like $\xi_0(p)$ curve agrees in its
behaviour well with the recent experimental result [25] on the
whole doping scale. One can state also some agreement with the
different $\xi_{ab}(p)$ behaviour in different fragmental regions
given  in [20-23], however these data cannot be joined
continuously with our curve.

Theoretical doping dependence of the c-axis second critical field
($T=0$) $H_{c2}=\Phi (2\pi\xi_0\sp 2)\sp{-1}$ is given in Fig.2.
This curve shows an intensive maximum built up from $c_{\beta}$
until $T_c$ maximum is reached.

In this respect there is also the agreement with the paper [25],
however the $H_{c2}$ maximum in [25] lies nearly at the doping
level which corresponds to the maximum of the collective pinning
energy. Our theoretical $H_{c2}$ peak corresponds to the effective
enhancement of the contribution of the cold subsystem with the
inside shifted chemical potential. This defect subsystem possesses
the smaller superconducting gap and supports larger $\xi_{ab}$.
Note that in the present model  $\Delta_1<\Delta_2$ if
$\rho_1>\rho_2$.

%\begin{figure}
%\resizebox{0.5\columnwidth}{!}

%\vspace{5cm}       % Give the correct figure height in cm
%\caption{The c-axis second critical magnetic field vs. doping}
%\end{figure}

The thermodynamic critical field
\begin{equation}
H_c(0)=[4\pi [\rho_{\gamma}\Delta_{\gamma}\sp
2(0)+(\rho_{\alpha}+\rho_{\beta}) \Delta_{\alpha}\sp
2(0)]]\sp{1/2}
\end{equation}
characterizes the condensation energy and is shown vs doping in
Fig.3. This is also a curve with a maximum (as observed [31])
which repeats the behaviour of $T_c$, of the superconducting
density $n_s$ [11,12] and the superconducting gaps. This means
that in the present model the strength of the pairing and the
phase coherence develop and vanish simultaneously. Such behaviour
of cuprates is stressed as the result of recent experimental
investigations [13-18]. The parallel course of the condensation
energy and of the superconducting gap has been followed in [32].
It must be mentioned that the present model delivers a sublinear
Uemura-type limited segment in the $T_c$ vs $n_s$ plot at
underdoping [12] with a followed back-turn along the branch
characterizing the overdoped side.

%\begin{figure}
%\resizebox{0.5\columnwidth}{!}

%\vspace{5cm}       % Give the correct figure height in cm
%\caption{The thermodynamic critical magnetic field on the doping
%scale.}
%\label{fig:3}       % Give a unique label
%\end{figure}

In conclusion, the present simple model, which describes
qualitatively correctly the behaviour of the cuprate energetic
characteristics, serves the same result for coherence properties
in agreement with the recent experimental data on the whole doping
scale. The qualitative agreement between the recent experimental data
 and the present theoretical approach on the nature of the cuprate
coherence length doping dependence seems to be of some importance,
at least for the reliability of the model.

This work was supported by the Estonian Science Foundation grant
No 6540.

\section*{References}
\begin{itemize}
\item[[1]] H. Suhl, B.T. Matthias, L.R.Walker, Phys. Rev. Lett.
{\bf 3} (1959) 552.
\item[[2]] V.A.  Moskalenko, Fiz. Met. Metalloved. {\bf 8} (1959) 503.
\item[[3]] A.Y. Liu, I.I. Mazin, J. Kortus, Phys. Rev. Lett. {\bf 87} (2001)
087005.
\item[[4]] N. Kristoffel, T. \"Ord, K. R\"ago, Europhys. Lett.
{\bf 61} (2003) 109.
\item[[5]] I.I. Mazin, V.P. Antropov, Physica C {\bf 385}, 49 (2003).
\item[[6]] N. Plakida, High-Temperature Superconductivity, \newline Springer,
Berlin, 1995.
\item[[7]] N. Kristoffel, P. Konsin, T. \"Ord, Rivista Nuovo Cim.
{\bf 17} (1994) 1.
\item[[8]] N. Kristoffel, P. Rubin, Physica C {\bf 356} (2001) 171.
\item[[9]] N. Kristoffel, P. Rubin, Eur. Phys. J. B {\bf 30} (2002) 495.
\item[[10]] N. Kristoffel, P. Rubin, Solid State Commun. {\bf 122} (2002) 265.
\item[[11]] N. Kristoffel, P. Rubin, Physica C {\bf 402}, 257 (2004).
\item[[12]] N. Kristoffel, P. Rubin, J. Supercond. (in press).
\item[[13]] C. Bernhard et al., Phys. Rev. Lett. {\bf 86} (2001) 1614.
\item[[14]] D.L. Feng et al., Science {\bf 289} (2000) 277.
\item[[15]] J.L. Tallon et al., Phys. Rev. B {\bf 68} (2003) 180501(R).
\item[[16]] T. Schneider, Physica B {\bf 326} (2003) 289.
\item[[17]] M.T. Trunin, Yu. Nefyodov, A.F. Shevchuk, Phys. Rev. Lett.
{\bf 92} (2004) 067006.
\item[[18]] R.H. He et al., Phys. Rev. B {\bf 69} (2004) 220502(R).
\item[[19]] H. Ihara, Physica C {\bf 364-365} (2001) 289.
\item[[20]] Y. Ando, K. Segava, Phys. Rev. Lett. {\bf 88} (2002) 167005.
\item[[21]] S.R. Curr\'as {\it et al.}, Phys. Rev. B {\bf 68} (2003) 094501.
\item[[22]] Y. Wang {\it et al.}, Science {\bf 299} (2003) 86.
\item[[23]] N.P. Ong, Y. Wang, Physica C {\bf 408-410} (2004) 11.
\item[[24]] M.A. Kastner {\it et al.}, Rev. Mod. Phys. {\bf 70} (1998) 897.
\item[[25]] H.H. Wen {\it et al.}, Europhys. Lett. {\bf 64} (2003) 790.
\item[[26]] P. Konsin, T. \"Ord, Physica C {\bf 191} (1992) 469.
\item[[27]] T. \"Ord, N. Kristoffel, Physica C {\bf 331} (2000) 13.
\item[[28]] H.-D. Dong Shen {\it et al.}, Phys. Rev. Lett. {\bf 93} (2004) 187002.
\item[[29]] A. Fang {\it et al.}, Phys. Rev. B {\bf 70} (2004) 214514.
\item[[30]] K.M. Lang {\it et al.}, Nature {\bf 415} (2002) 412.
\item[[31]] J.W. Loram {\it et al.}, Physica C {\bf 282-287} (1997) 1405.
\item[[32]] T. Shibauchi {\it et al.}, Phys. Rev. Lett. {\bf 86} (2001) 5763.
\end{itemize}

\newpage

Figure captions.

Fig.~1  Cuprate transition temperature (dashed line) and the
ab-plane coherence length vs hole doping.\\

Fig.~2  The c-axis second   critical magnetic field vs. doping.\\

Fig.~3  The thermodynamic critical magnetic field on the doping
scale (eV$^{1/2}$).

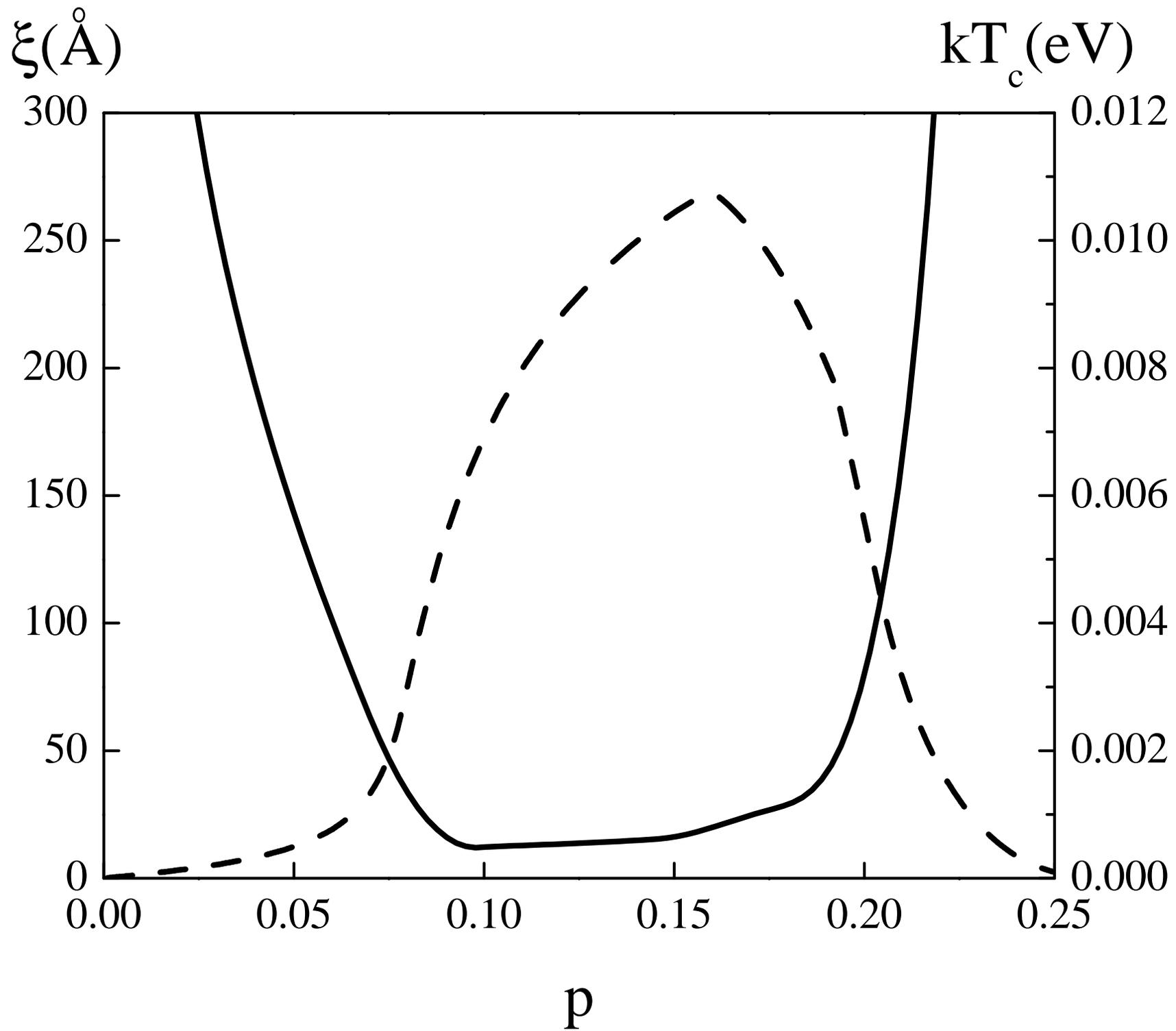
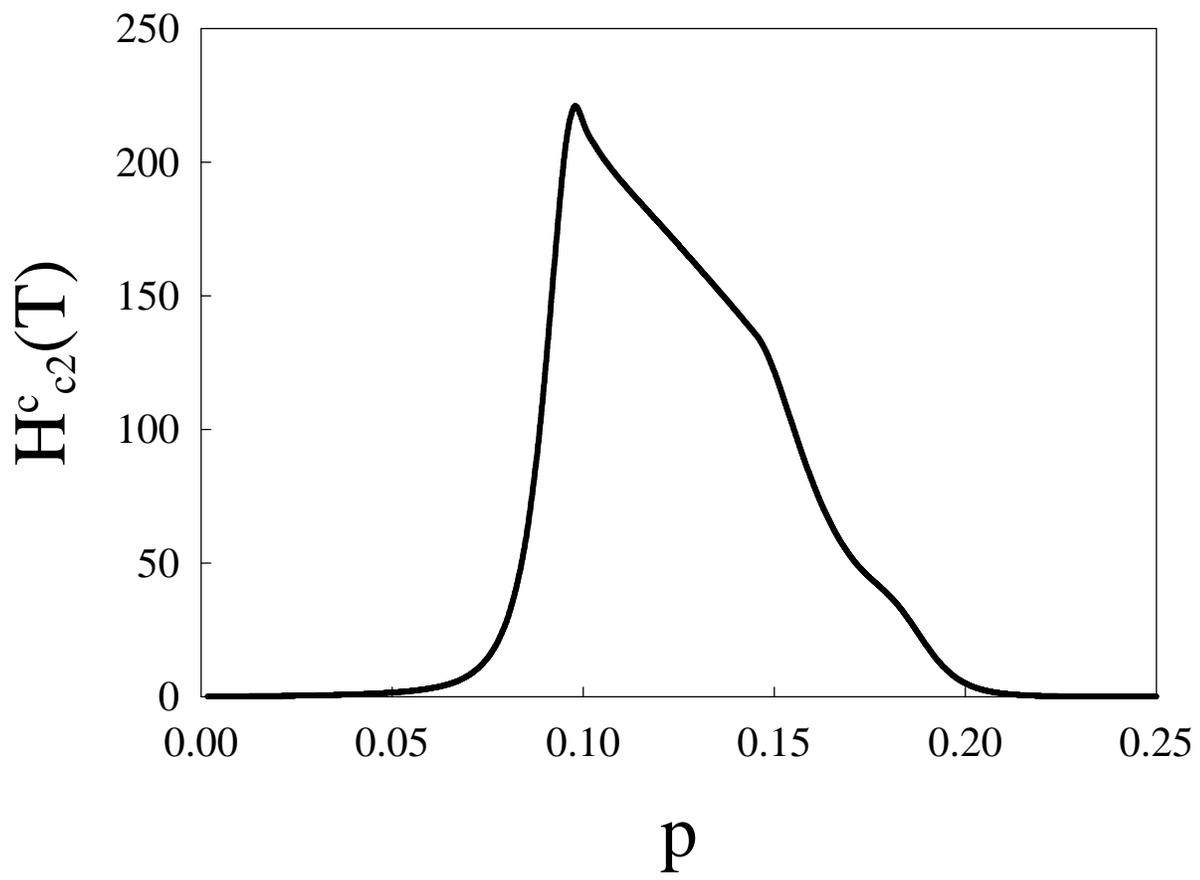
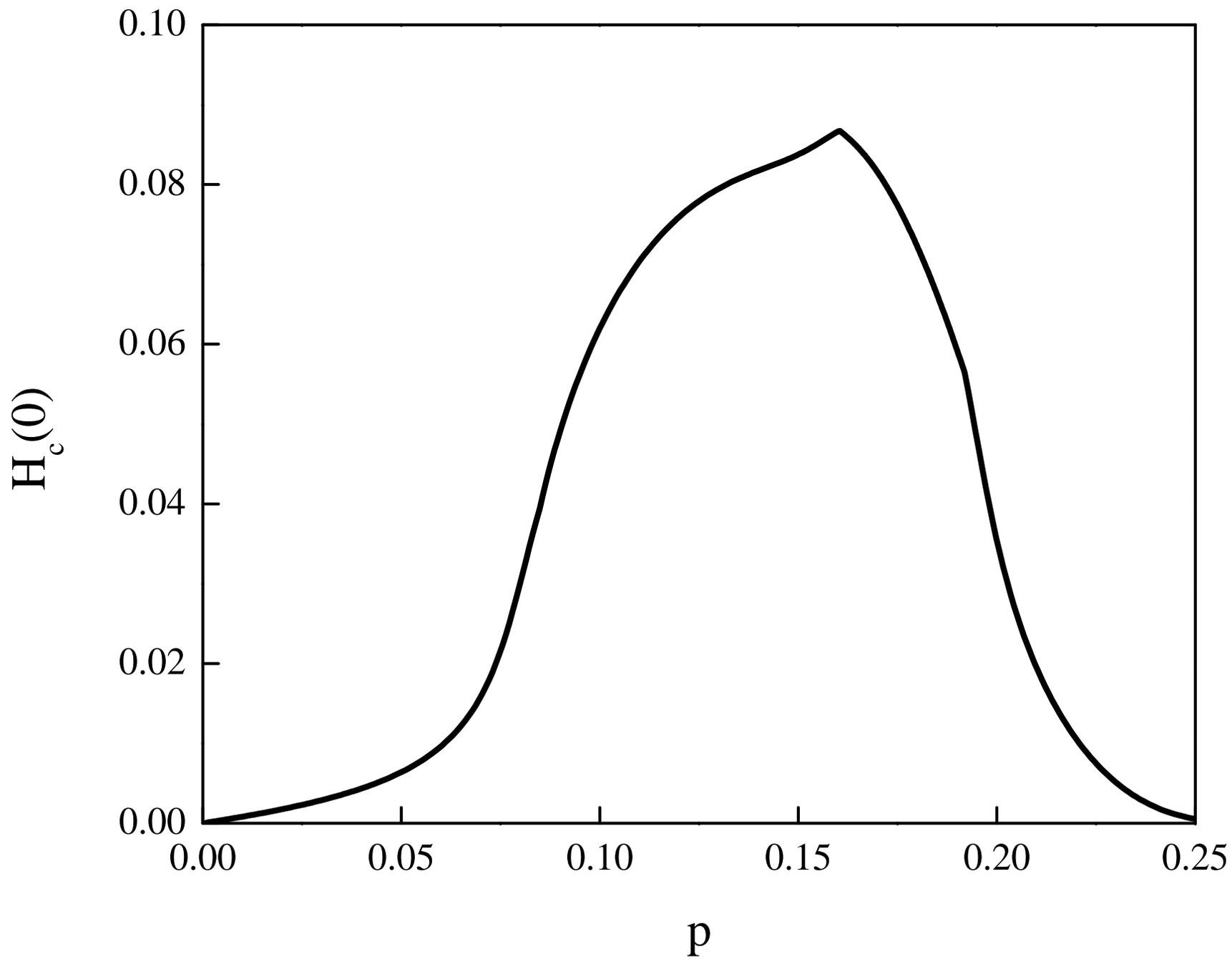
\end{document}